\documentclass[
    twocolumn,
    10pt,
    aps,
    pra,
    superscriptaddress,
    floatfix,
    citeautoscript
]{revtex4-2}
\usepackage[T1]{fontenc}
\usepackage[dvipsnames]{xcolor}
\usepackage{amsthm}
\usepackage{thmtools}
\usepackage{mathtools}
\usepackage{braket}
\usepackage{bm} 
\usepackage{bbm} 
\usepackage{dsfont}
\usepackage{amsmath}
\usepackage{amssymb}
\usepackage{enumitem}
\usepackage{comment}
\usepackage{float}
\usepackage[colorlinks]{hyperref}
\hypersetup{
    breaklinks  = true,
    colorlinks  = true,
    citecolor   = PineGreen,
    linkcolor   = MidnightBlue,
    urlcolor    = PineGreen
}
\usepackage{inputenc}
\usepackage{tikz}
\usepackage{pgfplots}
\usepackage[compat=0.4]{yquant}
\usepackage{xurl}
\usepackage{cleveref}

\usepackage{graphicx}
\usepackage{caption}

\usepackage{etoolbox}
\apptocmd{\sloppy}{\hbadness 10000\relax}{}{}

\usepackage{silence}
\usepackage{amsfonts,dsfont}
\usepackage{mathrsfs}



\usepackage[colorlinks]{hyperref}
\hypersetup{
    colorlinks  = true,
    citecolor   = PineGreen,
    linkcolor   = MidnightBlue,
    urlcolor    = PineGreen
}

\definecolor{fuchsia}{rgb}{0.96, 0.0, 0.63}
\definecolor{dur}{rgb}{0.6, 0, 0.1}

\definecolor{gnuplot_green}{HTML}{009E73}
\definecolor{gnuplot_purple}{HTML}{9400D3}

\AtBeginDocument{}
\AtBeginDocument{}
\pgfplotsset{compat=1.18}
\begin{document}

\title{Advantage of Warm Starts for Electron-Phonon Systems on Quantum Computers}

\date{\today}

\author{Arnab \surname{Adhikary}}
\email{arnab@phas.ubc.ca}
\affiliation{Department of Physics and Astronomy, University of British Columbia, 
Vancouver, BC V6T 1Z1, Canada}
\affiliation{Stewart Blusson Quantum Matter Institute, University of British Columbia, Vancouver, BC V6T 1Z4, Canada}
\affiliation{Institut f\"ur Theoretische Physik, Leibniz Universit{\"a}t Hannover, Appelstr. 2, 30167 Hannover, Germany}

\author{S. E. \surname{Skelton}}
\email{shawn.skelton@itp.uni-hannover.de}
\affiliation{Institut f\"ur Theoretische Physik, Leibniz Universit{\"a}t Hannover, Appelstr. 2, 30167 Hannover, Germany}

\author{Alberto  \surname{Nocera}}
\email{berciu@phas.ubc.ca}
\affiliation{Department of Physics and Astronomy, University of British Columbia, 
Vancouver, BC V6T 1Z1, Canada}
\affiliation{Stewart Blusson Quantum Matter Institute, University of British Columbia, Vancouver, BC V6T 1Z4, Canada}

\author{Mona  \surname{Berciu}}
\email{berciu@phas.ubc.ca}
\affiliation{Department of Physics and Astronomy, University of British Columbia, 
Vancouver, BC V6T 1Z1, Canada}
\affiliation{Stewart Blusson Quantum Matter Institute, University of British Columbia, Vancouver, BC V6T 1Z4, Canada}

\begin{abstract}

Simulating electron–phonon interactions on quantum computers remains challenging, with most algorithmic effort focused on Hamiltonian simulation and circuit optimization. In this work, we study the single-electron Holstein model and propose an initial-state ansatz that substantially enhances ground-state overlap in the strong-coupling regime, thereby reducing the number of iterations required in standard quantum phase estimation. We further show that this ansatz can be implemented efficiently and yields an exponential reduction in overall circuit costs relative to conventional initial guesses. Our results highlight the practical value of incorporating physical intuition into initial state preparation for electron–phonon coupled systems.

\end{abstract}

\maketitle

\section{Introduction}
\label{sec:introduction}

Accurately describing correlated fermions, bosons and their interactions remains one of the central challenges in chemistry and condensed matter physics. Quantum computers provide a fundamentally new way to simulate quantum many-body systems whose complexity exceeds the reach of classical computation.  Quantum algorithms can, in principle, represent these exponentially large Hilbert spaces using only polynomial resources, offering direct access to ground and excited states that are inaccessible to classical methods~\cite{Feynman1982,Lloyd1996, Abrams1999}.

A key step in many such simulations is the preparation of a high-fidelity low-energy initial state, for example in workflows leveraging quantum phase estimation (QPE) to compute the ground state of a given Hamiltonian~\cite{Kitaev1995,AspuruGuzik2005}. The efficiency of QPE depends critically on the overlap between the prepared state and the true ground state: when this overlap is small, the number of repetitions and overall runtime increase rapidly. For weakly correlated systems, mean-field product states such as Hartree–Fock often provide an adequate starting point, but in strongly correlated regimes these approximations fail to capture the essential entanglement and collective structure of the ground state.

Electron–phonon systems present a particularly rich and demanding setting for quantum simulation. In these systems, itinerant electrons interact with quantized lattice vibrations, giving rise to a range of phenomena including  polaron and bipolaron formation, superconductivity, charge density waves.  Their combined fermionic and bosonic character, however, makes them exceptionally difficult to simulate on classical hardware. A comprehensive quantum computing framework for such systems was developed in Ref.~\cite{Macridin_PRA, Macridin_PRL}, demonstrating how to encode electron–phonon Hamiltonians such as the Hubbard–Holstein~\cite{Holstein1959} and Su–Schrieffer–Heeger (SSH) models~\cite{SSH1979}, and how to use QPE to extract their energy spectra starting from an initial uncorrelated state describing free electrons. This work established formally the feasibility of simulating correlated electron–phonon dynamics with polynomial resources.

In practice, however, the efficiency of these simulations is severely limited in the strong coupling regime. Here the ground state consists of electrons that are heavily dressed by clouds of phonons describing the lattice distortions created by electrons in their vicinity, forming entangled polaronic states~\cite{LangFirsov1963}. Even in the single electron limit, the overlap between the true ground-state and a non-interacting electron initial state becomes exponentially small at strong coupling, leading to high sampling overheads for QPE. In this regime, the computational cost is determined not by the cost of Hamiltonian simulation but by the fidelity of the initial state.

The present work begins to address this limitation by focusing on an efficient and physically motivated initial state preparation  for the study of the single Holstein polaron. The structure of the ground state phonon cloud is well understood in the limit of very strong coupling, through transformations such as Lang–Firsov that incorporate electron–phonon dressing at the operator level~\cite{LangFirsov1963}. Building on this theoretical insight, we construct an initial state consistent with the known strong-coupling form and develop a compact quantum circuit that prepares it efficiently. The circuit exploits symmetries and factorization properties of the model to minimize depth and gate count, enabling scalable implementation.

We analyze the resource requirements of this approach and show that the cost of preparing this non-trivial initial state is negligible compared to the total runtime of energy estimation. At the same time, the improved overlap with the true ground state significantly reduces the number of QPE iterations required to reach a given energy precision. Thus,  this work demonstrates that incorporating well-established theoretical insights in the choice of the initial state can yield substantial algorithmic advantages when realized as efficient quantum circuits. Applied to the Holstein model, this approach provides a concrete path to overcoming the strong-coupling bottleneck in quantum simulations of electron–phonon systems and, more generally, to improving the scalability of quantum algorithms for correlated systems.

The remainder of the paper is organized as follows. Section~\ref{sec:preliminaries} introduces the physical model and outlines the ground-state preparation problem. Section~\ref{sec:ansatz} describes the proposed initial state and the physical rationale behind it. Section~\ref{sec:circuitimplementation} presents the corresponding quantum circuit. Section~\ref{sec:resourcecost} analyzes resource requirements, with emphasis on circuit depth. Section~\ref{sec:conclusion} discusses the broader implications of this work and possible future directions. Full state preparation schemes  and other technical details are presented in the Appendixes.

\section{Background and Setting}
\label{sec:preliminaries}

The Holstein model \cite{Holstein1959,Holstein1959b} provides a minimal setting to study the interplay between electronic motion and local lattice vibrations. For a single polaron (when electron-electron correlations cannot arise), the  Hamiltonian is written as the sum of three contributions.
\begin{equation}
H = H_e + H_p + H_{e-ph}.
\end{equation}
The electronic part is given by
\begin{equation}
H_e = -t \sum_{\langle i,j \rangle} c_i^\dagger c_j ,
\end{equation}
describing nearest-neighbor hopping  with amplitude $t$ of the electron. The phonons are typically modeled as independent harmonic oscillators of frequency $\omega_0$ located on each lattice site (we set $\hbar =1$ and ignore the constant zero-point energy),
\begin{equation}
H_{ph} = \omega_0 \sum_i b_i^\dagger b_i ,
\end{equation}
where $b_i^\dagger$ ($b_i$) creates (annihilates) a phonon at site $i$. Finally, the Holstein electron--phonon coupling is local,
\begin{equation}
H_{e-ph} = g \sum_i n_i \left(b_i^\dagger + b_i\right) ,
\end{equation}
where $n_i = c_i^\dagger c_i$ is the electron number operator and $g$ is the coupling strength.

In this formulation, $H_e$ promotes delocalization of the electron, while $H_{ep}$ favors localization by binding the electron to a local lattice distortion. The balance between these competing effects determines the qualitative behavior of the system. In the weak-coupling regime ($g \ll t, \omega_0 $), the ground state resembles a nearly free electron weakly dressed by a small number of phonons that can be located spatially quite far from the electron -- a so-called \emph{large polaron}. In contrast, when the coupling becomes comparable to or larger than the geometric mean of the electronic bandwidth and the phonon energy such that the effective coupling $\lambda = g^2/(2t \omega_0)\gtrsim 1$, the ground-state crosses over to a \emph{small polaron}. This quasiparticle also consists of the electron and its associated phonon cloud,  which however now has many phonons located very close to the electron. Because of carrying around this significant cloud, the polaron acquires an exponentially increased effective mass, and its motion is strongly suppressed.

The structure of the ground state in the single-site limit $t=0$ (formally equivalent with infinite effective coupling $\lambda \rightarrow \infty$) illustrates the challenge for quantum algorithms. If we set $t=0$, the electron is localized at some site $i$ and only interacts with that site's phonons. The corresponding  ground-state can be found straightforwardly
\begin{equation}
|i\rangle_E 
\otimes  |-\alpha\rangle_{B_i}= c^\dagger_i |{\rm vac}\rangle_E \otimes  e^{-{\alpha^2\over 2} - \alpha b^\dagger_i}|{\rm vac}\rangle_{B}
\label{LF}
\end{equation}
where $|{\rm vac}\rangle_E, |{\rm vac}\rangle_B$ are the vacuums for electrons and phonons, respectively. The phonon coherent state shows that the oscillator at site $i$ is displaced out of equilibrium by
\begin{equation}
\alpha = g / \omega_0,
\end{equation}
{\em i.e. } the local lattice  distortion increases linearly with the electron-phonon coupling $g$.

The overlap between this $t=0$ ground-state and an initial state $c_i^\dagger|{\rm vac}\rangle_E\otimes|{\rm vac}\rangle_{B} $  describing a localized and undressed electron (no phonon cloud) is thus exponentially suppressed by the phonon contribution to the overlap:
\begin{equation}
_B\braket{{\rm vac} |-\alpha}_{B_i} = e^{-\alpha^2/2}.
\label{free}
\end{equation}
Turning on a finite $t$ restores the invariance to translations so that momentum becomes a good quantum number and the polaron is equally likely to be at any site. Nevertheless, at strong couplings where the polaron moves very slowly,  the phonon cloud associated with each electron position resembles that of Eq. (\ref{LF}). This is why at strong couplings, the overlap between the true ground-state and that describing a free electron with zero momentum decreases exponentially with $\lambda$.

\subsection*{Quantum Phase Estimation and the Role of Initial State Overlap}

QPE is a central algorithmic primitive for obtaining eigenvalues of a Hamiltonian $H$ with high precision. Given access to controlled time-evolution $e^{-iHt}$, the algorithm extracts eigenphases through interference on an ancillary register, ultimately producing an estimate of an eigenenergy $E_j$ while projecting the system register onto the corresponding eigenstate $\ket{E_j}$. The probability of obtaining a particular outcome is determined by the squared overlap between the initial state $\ket{\psi_{\text{init}}}$ and that eigenstate,
\begin{equation}
    \Omega_j = |\braket{E_j | \psi_{\text{init}}}|^2 .
\end{equation}

In the context of ground-state energy estimation, the quantity of interest is the overlap $\Omega_{\text{gs}}=|\braket{\rm GS | \psi_{\text{init}}}|^2$, where $|\rm GS\rangle$ is the true ground-state wavefunction. If $\Omega_{\text{gs}}$ is large, a single execution of QPE suffices with high probability. Conversely, if $\Omega_{\text{gs}}$ is small, the likelihood of projecting onto the ground state diminishes, and the expected number of repetitions grows as $\mathcal{O}(1/\Omega_{\text{gs}})$. Although amplitude amplification techniques can in principle boost the success probability, they typically incur an additional cost scaling as $\mathcal{O}(1/\sqrt{\Omega_{\text{gs}}})$ in controlled applications of $e^{-iHt}$. As a result, the efficiency of QPE is governed not only by the cost of Hamiltonian simulation, but also, critically, by the choice of initial state.

This explains the issue with choosing a phonon-free initial state for an electron--phonon system in the strong-coupling regime. As discussed above, its overlap with the true polaron ground-state decreases exponentially with the coupling strength in this regime. As a result, the asymptotic scaling advantages of QPE are rendered irrelevant: an exponentially small $\Omega_{\text{gs}}$ implies an exponentially large number of repetitions to achieve constant success probability. This phenomenon is closely related to the ``orthogonality catastrophe,'' where correlations in the true ground state make it nearly orthogonal to simple reference states even in finite-size systems.

To summarize, while Hamiltonian simulation and bosonic encodings determine the per-run cost of QPE, the initial state overlap determines how many runs are required. For practical quantum algorithms, preparing a physically motivated initial state with substantial ground-state support is essential. In the following, we introduce such an ansatz for the Holstein polaron, inspired by the Lang--Firsov $t=0$ solution discussed above, and show that it leads to significant improvement in the ground state overlap $\Omega_{\text{gs}}$.

\section{Lang-Firsov ansatz and its overlap with the true ground-state}
\label{sec:ansatz}

We propose to use as a better initial state in the strong-coupling limit, the wavefunction:
\begin{align}
    |\Psi_{\text{LF}}(\alpha)\rangle= \frac{1}{\sqrt{N}}\sum_{i=1}^{N} c_i^{\dagger} e^{-\frac{\alpha^2}{2}-\alpha b_i^{\dagger}} \, |0 \rangle
\label{ini}
\end{align}
where $N$ is the number of sites in the system. Note that if we set $\alpha=0$, this reduces to the phonon-free initial state used in Ref. \cite{Macridin_PRL,Macridin_PRA}. If we set $\alpha = g/\omega_0$, this is just the prediction of first-order perturbation theory in $t$, starting from the $t=0$ ({\em i.e.} strong coupling) limit. More generally, one can let $\alpha$ be a variational parameter and choose the value that minimizes the total energy. In this case, $\alpha$ is given by the implicit equation
\begin{equation}
\alpha = {g\over \omega_0} - {2t \alpha\over \omega_0} e^{-\alpha^2}.
\label{var}
\end{equation}

Next, we rewrite Eq. (\ref{ini}) in a form that makes its structure transparent. 
In general, simulating fermionic systems on qubits requires mappings such as the 
Jordan--Wigner transformation to enforce fermionic anticommutation relations. 
However, in the single-electron limit these sign structures never appear, 
and the mapping is trivial: the electronic Hilbert space is simply the span 
of $\{\,|i\rangle_E\,\}_{i=1}^N$, where 
\(|i\rangle_E \equiv c_i^\dagger |{\text{vac}}\rangle_E\)
denotes an electron localized at site \(i\);  $|{\text{vac}}\rangle_E$ is the electron vacuum.

The factor 
\(\exp[-\frac{\alpha^2}{2} - \alpha b_i^\dagger]\)
is the normally ordered form of a displacement operator acting on the phonon vacuum, producing the coherent phonon state 
\(|-\alpha\rangle_{B_i}\) at site $i$, while
all other oscillators remain in their vacuum states
\(|\text{vac}\rangle_{B_j}\) for \(j \neq i\).
Thus each term in the sum in Eq. (\ref{ini}) corresponds to an electron at site \(i\),
a displaced phonon cloud at that site,
and vacuum phonons elsewhere.

Collecting these contributions yields
\begin{equation}
|\Psi_{\mathrm{LF}}(\alpha)\rangle 
= \frac{1}{\sqrt{N}} 
\sum_{i=1}^N 
|i\rangle_E 
\otimes 
|-\alpha\rangle_{B_i} 
\bigotimes_{j\neq i} |\text{vac}\rangle_{B_j},
\end{equation}
which shows explicitly that the Lang--Firsov state is a delocalized superposition of electron--phonon product states,
with the electron at each site dressed by a coherent local phonon displacement.

\begin{figure}[t]
    \includegraphics[width=1.05\linewidth]{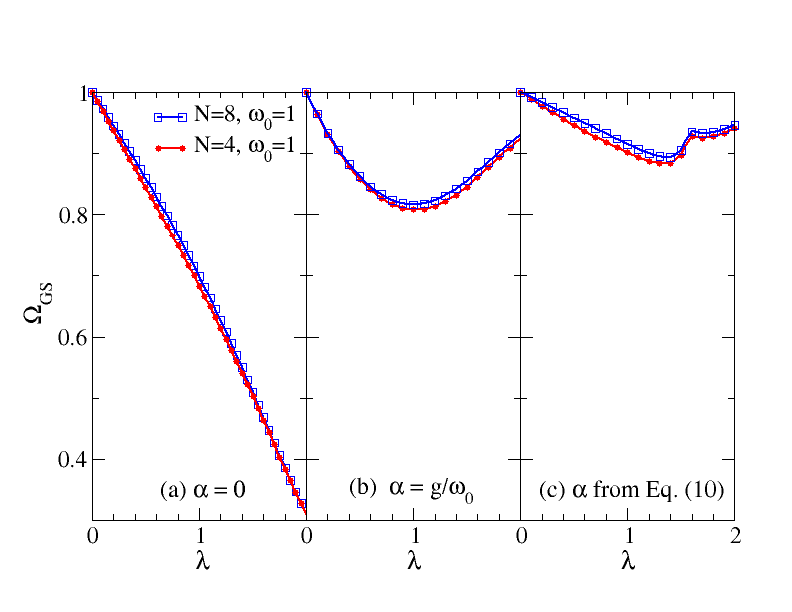}
\caption{\raggedright  Overlap $\Omega_{\rm gs}=|\langle {\rm GS}|\Psi_{\text{LF}}(\alpha)\rangle|^2$ versus $\lambda$ for $t=\omega_0=1$ and (a) the phonon-free initial state
with $\alpha = 0$; (b) the strong-coupling value $\alpha = g/\omega_0$; and
(c) the variational solution of Eq. (10). The exact GS was calculated with DMRG for chains of various lengths $N$. We only show results for $N=4$ and $N=8$ because values for $N>8$ are indistinguishable from those with $N=8$, on this scale.
}
    \label{fig1}
\end{figure}

 As already mentioned, for $N=1$, $|\Psi_{\mathrm{LF}}(\alpha = g/\omega_0)\rangle$ is the actual ground state so its $\Omega_{\text{gs}}=1$ for all couplings $g$. The case $N=2$ can also be solved (semi) analytically \cite{berciuexact2007} but the value for the corresponding $\Omega_{\text{gs}}$ is not available in a simple form. We use density matrix renormalization group (DMRG) to calculate the polaron ground-state for chains of various length $N$, and to then obtain their corresponding overlaps $\Omega_{\text{gs}}$ with $|\Psi_{\text{LF}}(\alpha)\rangle$. Fig. \ref{fig1} shows typical results obtained for $\omega_0=1$. The panels show the overlap $\Omega_{\rm gs}=|\langle {\rm GS}|\Psi_{\text{LF}}(\alpha)\rangle|^2$ versus $\lambda$ for (a) the phonon-free initial state with $\alpha=0$; (b) the strong-coupling value $\alpha=g/\omega_0$; and (c) the variational solution of Eq. (\ref{var}). Clearly, both initial states with finite $\alpha$ have significantly higher overlaps $\Omega_{\rm gs}$ compared to the phonon-free initial state with $\alpha=0$, especially at larger $\lambda$. In particular, the variational solution of Eq. ({\ref{var}) has $\Omega_{\rm gs}>0.9$ for (nearly) the entire range of possible couplings, proving to be an excellent initial guess at all coupling strengths. Only values for chains with $N=4,8$ sites are shown in Fig. \ref{fig1} because results for $N>8$ are indistinguishable from those with $N=8$, on this scale.

It is worth emphasizing that for the parameters used in Fig. \ref{fig1}, the system is not yet in the strongly coupled limit even at $\lambda =2$; even here, the overlap with the zero-phonon state is $\sim 0.3$,  so implementing the finite-$\alpha$ initial state might not seem to offer a significant speed-up. However,  for even larger $\lambda$ and/or smaller values of $\omega_0$, and in higher dimensions, the crossover to the strong coupling limit occurs at smaller values $\lambda\approx 1$ and is signaled by the overlap $\Omega_{\rm gs}$ becoming exponentially small, see Eq. (\ref{free}) (specific results are shown, for {\em e.g.}, in Ref. \cite{GoodvinPRB2006}). By contrast, in the strong-coupling limit it is guaranteed that  $\Omega_{\rm gs}\rightarrow 1$ for both finite values of $\alpha$,  so using the initial state with finite $\alpha$ will provide a very significant reduction in the number of QPE repetitions.

\section{Quantum Circuit Implementation}
\label{sec:circuitimplementation}

To prepare the proposed initial state ansatz on a quantum computer, we begin by mapping the electronic degrees of freedom onto qubit registers. 
The electronic registers specify the lattice site occupied by the electron. 
To encode this information, we require 
$k_E = \lceil \log_2 N \rceil $  qubits, where \( N \) is the number of sites. Without loss of generality we will assume herein that $\log_2 N $ is an integer.
Each computational basis state represents a binary encoding of a site index. 
Applying Hadamard gates to all qubits produces an equal superposition over all basis states, 
reflecting the translational invariance of the system—physically, the electron has equal probability of being located at any site.

We now focus on the phononic degrees of freedom. The infinite-dimensional phonon Hilbert space must be truncated. We follow the approach in Ref. \cite{Macridin_PRL} and allocate 
$m$ qubits to represent each local phonon mode, with 
$m\approx 6-8$ typically sufficient to capture the low-lying phononic subspace of interest. Entangling operations between the fermionic and bosonic qubits implement the electron–phonon coupling, thereby encoding the relevant correlations in the system.

\begin{figure}
\centering
\begin{tikzpicture}
\begin{yquant}
qubit {$\ket{j_{\idx}}_E=\ket{0}_E$} j[2];
qubits {$\ket{0}_B$} b[4];
box {$U_0$} b;
H j;
box {$\tilde{U}(\alpha)$} (b[0]) | ~ j;
box {$\tilde{U}(\alpha)$} (b[1]) |  j[1], ~j[0];
box {$\tilde{U}(\alpha)$} (b[2]) |  j[0], ~j[1];
box {$\tilde{U}(\alpha)$} (b[3]) | j;
\end{yquant}
\end{tikzpicture}

\caption{\raggedright Circuit for the Lang–Firsov ansatz on a chain of length \(N=4\). The first two electron registers specify the electron’s position. Each of the four remaining registers, with \(m \approx 6\!-\!8\) qubits per register, encodes the local phonon mode at its site. \(U_{0}\) prepares the harmonic–oscillator ground state on every phonon register. The controlled–\(\tilde U(\alpha)\) applies the site-dependent phonon displacement conditioned on the electron registers.}
\label{fig:lang-firsov-circuit}
\end{figure}
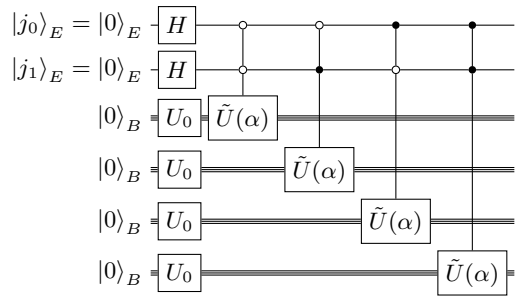

Conceptually, the circuit consists of two main components. 
The first unitary, \( U_0 \), acts on the qubits corresponding to a local phonon mode and prepares the vacuum (ground) state 
of the harmonic oscillator within the truncated Hilbert space. For any site, starting from the computational basis state 
\(|0\rangle_B \equiv |00\cdots0\rangle\) of the \(m\)-qubit register, the transformation
\[
U_0 |0\rangle = |\text{vac}\rangle_B
\]
maps the digital zero state of the qubits to the encoded oscillator vacuum state \(|\text{vac}\rangle_B\), which approximates 
the ground state of that site's bosonic mode. 

The second unitary, \( \tilde{U}_\alpha \), acts as a displacement operator in the position basis, shifting the vacuum state 
prepared by \( U_0 \) to generate a coherent state of amplitude \(-\alpha\) at that site:
\[
\tilde{U}_\alpha U_0 |0\rangle_B = D(-\alpha)|\text{vac}\rangle_B = |-\alpha\rangle_B
\]
where
\[
\quad D(\alpha) = e^{\alpha a^\dagger - \alpha^* a}.
\]
This operation reproduces the action of the bosonic displacement operator on the encoded vacuum, 
creating a localized phonon excitation with amplitude \(\alpha\) at a site of interest.

In the complete circuit, 
\( \tilde{U}_\alpha \) must be applied conditionally only at the site where the electronic register indicates 
the presence of an electron. The control mechanism, realized through entangling gates between electronic and phononic 
qubits, ensures that the local phonon displacement occurs exclusively in the presence of the electron, so as to encode the initial state of Eq. (\ref{ini}).

Together, these two operations implement the Lang--Firsov ansatz on the quantum circuit: the electron locally displaces 
the phonon field, forming a correlated electron--phonon state that captures the essence of the polaronic coupling.

With the overall circuit structure established, we now describe the implementation of its two key components: 
the gaussian preparation unitary \( U_0 \) and the controlled displacement operation \( \tilde{U}_\alpha \).

The role of \( U_0 \) is to generate an approximate Gaussian wavefunction on the local phonon register of \( m \) qubits,  thereby preparing an encoded version of the oscillator vacuum state within the truncated Hilbert space.  The subsequent operation \( \tilde{U}_\alpha \) introduces electron-conditioned phonon displacements, realizing the electron–phonon correlations central to the Lang–Firsov ansatz.

The Gaussian state can be prepared using quantum eigenvalue transformation (QET). One begins from a block-encoding of $\sum \sin\left(\frac{2x}{2^m}\right)\ket{x}\bra{x}$, where $x\in[-W, W]$, and then implements a degree-$d$ polynomial transformation $p_0: [0, \sin(1)]\rightarrow [-1, 1]$. $p_0^{(d)}(y)$ is defined such that
\begin{equation}
    \left|\left|p_0^{(d)}(y)-e^{-i\frac{W^2}{2}(\arcsin(y))^2}\right|\right|_{\infty}\leq \epsilon.
\end{equation}
Here $e^{-i\frac{W^2}{2}(\arcsin(y))^2}$ is precisely the desired function, with rescaled variable $\arcsin(y)=x/2W$.

The (unnormalized) resultant state is
\begin{align}
    U_{p_0^{(d)}}&\left[\text{BE}\left(\sum \sin\left(\frac{2x}{2^m}\right)\ket{x}\bra{x}\right)\right]\ket{0, 0^{\otimes m}}\nonumber\\
    &={\ket{0}\otimes p_0^{(d)}(x)\ket{x}}+\ket{\psi^{\perp}}
\end{align}
which is brought to $\mathcal{O}(1)$ success probability with quantum amplitude amplification and a small overhead depending on the norm of $p_0^{(d)}$ over the interval. 


The controlled displacement operation \( \tilde{U}_\alpha \) introduces the nontrivial coupling between electronic and phononic modes. 
The key observation guiding its construction is that the corresponding phase transformation is diagonal in the momentum basis. 
For real displacement amplitudes \(\alpha \in \mathbb{R}\),
\[
D(\alpha) = e^{\alpha a^\dagger - \alpha a} = e^{-i x_0 P}, \qquad x_0 = \sqrt{2}\,\alpha,
\]
which represents a translation in position space generated by the momentum operator \(P\). 
Because \(e^{-i x_0 P}\) acts diagonally in the momentum basis with eigenvalues \(e^{-i x_0 p}\), 
it can be implemented by transforming to momentum space, applying the appropriate phase factors, and transforming back. 
This motivates the \emph{QFT– diagonal gates –iQFT} construction.

In practice, the local phonon displacement unitary \(\tilde{U}_\alpha\) is implemented with a discrete quantum Fourier transform (QFT) on the \(m\)-qubit phonon register to map position to momentum representation; then a single-qubit diagonal rotations \(R_z(\phi_\ell(\alpha))\) encoding the phase \(\phi_\ell(\alpha)\) associated with each binary-weighted momentum component; and finally  an inverse Fourier transform (iQFT) to return to the position basis.

The quantum circuit implementation is given in \cref{fig:utildecircuit} in the Appendix. All dependence on the displacement amplitude \(\alpha\) enters solely through the angles 
of these diagonal single-qubit rotations in the momentum-space representation. 
In this construction, the momentum-dependent shifts are encoded as local \(R_z(\phi_\ell(\alpha))\) operations acting independently on each qubit, 
making the role of \(\alpha\) conceptually transparent and practically isolated—only the diagonal phase gates depend on \(\alpha\), 
while all other elements of the state-preparation circuit remain unchanged.

To incorporate electron–phonon correlations consistent with the Lang–Firsov ansatz, 
the entire QFT–diagonal–iQFT sequence is controlled by the electronic register, 
ensuring that the displacement acts exclusively at the site occupied by the electron. 
This conditional operation reproduces the electron-induced local phonon deformation that defines the Lang–Firsov transformation.

\section{Resource Estimation}
\label{sec:resourcecost}
We quantify circuit cost using the number of $T$ gates, often referred to as ``magic'' gates. The $T$ count is a widely used proxy for quantum advantage, as $T$ gates separate classically simulable Clifford operations from an approximately universal gate set \cite{howardapplication2017, Veitch_2014, gottesman1998heisenbergrepresentationquantumcomputers}. For each of the two main subroutines shown in \Cref{fig:lang-firsov-circuit}, we first review an asymptotically optimal implementation strategy and then report the resulting $T$-count for our example, considering different truncations of the bosonic Hilbert space. All resource estimates are obtained using PennyLane \cite{bergholm2018pennylane} by decomposing the circuits into single- and two-qubit Clifford gates, Toffoli gates, and $T$ gates. We note that a full decomposition of Toffoli gates into Clifford+$T$ primitives would lead to a modest increase in the reported $T$-counts.

\subsection{Cost of preparing Harmonic Oscillator Ground State}
QET techniques can be used to prepare the harmonic oscillator ground state efficiently. Doing so consists of preparing an inexpensive block-encoding of $\sin(2x/2^m)$ for $x$ within discrete grid points $0,...2^m-1$, using QET to prepare a block-encoding of $e^{-\frac{x^{2}}{2}}$ to accuracy $\epsilon$. The QET circuit depth is directly controlled by the polynomial approximation degree $d$, which can usually be lower-bound with respect to the desired precision. Finally, quantum amplitude amplification is used to drive the algorithm to $\mathcal{O}(1)$ success probability. The circuit which $\epsilon$-approximates the desired ground state in the position basis representation scales favorably with the problem parameters \cite{mcardle2025quantumstatepreparationcoherent}. Overall, one expects to use $\mathcal{O}\left(m\log^{4/5}(\epsilon^{-1})\right)$ multi-qubit gates to prepare a Gaussian state.

In our implementation, accounting for the degree and sub-normalization of our approximation polynomial, we expect 
\begin{equation}
N_{mult-controlled}=\mathcal{O}\left(4\cdot 22\cdot m\right),\label{eq:qsvtcostscaling}
\end{equation}
where $N_{mult-controlled}$ is the cost of the multi-controlled gates, which dominates the overall cost scaling. The exact cost is synthesized from the circuits in \cref{fig:qsp-nonhermit-circuit-real} and \cref{fig:be-symm-circuit}. The circuit uses $m+3$ qubits.



\subsection{Cost of preparing the Lang-Firsov Ansatz}
Once an approximate ground-state wavefunction has been prepared, incorporating the Lang--Firsov ansatz adds only the cost of implementing the \emph{controlled displacement operator} that entangles the electron and phonon degrees of freedom.  
This displacement, \(e^{-i x_0 P}\), is realized using the standard change-of-basis construction,
\begin{equation}
e^{-i x_0 P} = \mathrm{QFT}^\dagger\, e^{-i x_0 X_{\mathrm{diag}}}\, \mathrm{QFT},
\end{equation}
where \(X_{\mathrm{diag}}\) is diagonal in the momentum basis.  
In the conditional version required for the single-electron case, only the \emph{phase-gradient block needs to be controlled}, since the surrounding QFT and inverse QFT cancel automatically whenever the control is inactive.

{For a phonon mode represented on \(m\) qubits, a QFT (or iQFT) requires exactly 
\(\tfrac{m(m-1)}{2}\) number of single qubit-controlled phase gates and \(m\) Hadamard gates.  
The intermediate phase-gradient step applies
$m$ number of single-qubit rotations \(R_z(\phi_j)\), one on each phonon qubit, each controlled by the same
$k_E = \lceil \log_2 N \rceil$-qubit electronic register.}

Hence, the full controlled-displacement layer acting on one phonon register requires
\begin{align}
N_{mult-controlled}&=\mathcal{O}\left(m^2\right),\label{eq:LF_ctrl}
\end{align}
Note that we could have approximated the Lang-Firsov displaced state with QET, just as was done for the ground state. In fact, from comparing \cref{eq:qsvtcostscaling} and \cref{eq:LF_ctrl}, we can see that for  $m^2\gtrsim m\log^{4/5}(1/\epsilon^{-1})$, it might become favorable to do so. However, the polynomial approximation degree and subnormalization then vary with the value of $\alpha$.

The cost of preparing the Lang-Firsov displaced state scales with the cost of multi-controlled rotation gates. As far as the cost of the controlled-$R_z$ gates (used to implement the displacement) is concerned, we first note that any single-qubit unitary \(U\) admits the decomposition $U=AXBXC,\text{ with } ABC = I,$ which reduces the implementation cost of a controlled-$U$ gate, $C^{k}(U)$, to that of two multi-controlled NOT $C^k(X)$ together with single-qubit gates.
For the remaining cost of the multi-controlled NOTs, there now exists a range of techniques spanning different resource priorities such as gate count~\cite{Gidney2015LargeCNOT, Nguyen2023} and circuit depth~\cite{Claudon2024PolylogCNOT}.



\subsection{Comparison of state preparation costs}
The full $T$ counts are provided in Fig.~\ref{fig:tcountfullcircuit}. Note that the $T$-count estimates are agnostic to the specific rotation angles, and so the magnitude of the displacement $\alpha$ does not factor into our analysis. Qualitatively, we observe that the $T$-count scales with the number of sites, as expected from the number of controlled $\tilde{U}({\alpha})$ steps. 

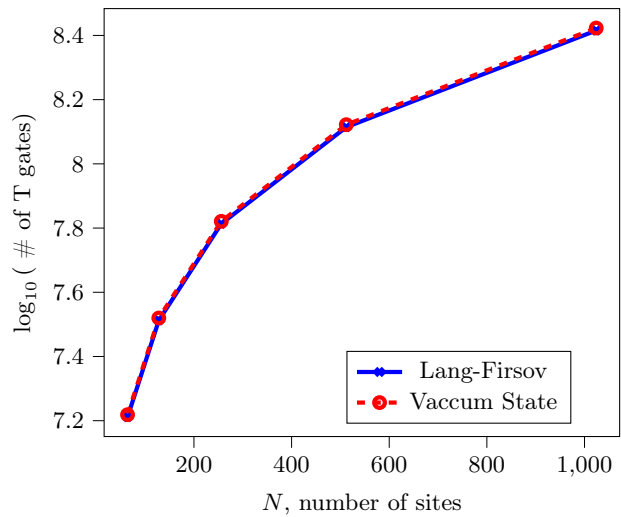
\begin{figure}[t]
    \centering
\begin{tikzpicture}

\definecolor{darkgray176}{RGB}{176,176,176}
\definecolor{darkorange25512714}{RGB}{255,127,14}
\definecolor{steelblue31119180}{RGB}{31,119,180}

\begin{axis}[
legend style={at={(0.47,0.03)},anchor=south west},
tick align=outside,
tick pos=left,
x grid style={darkgray176},
xlabel={$N$, number of sites},
ylabel={$\log_{10}\left(\text{ \#\ of T gates}\right)$ },
xmin=16, xmax=1072,
xtick style={color=black},
y grid style={darkgray176},
ymin=7.15067408402385, ymax=8.4835606061956,
ytick style={color=black}
]
\addplot [ultra thick, blue,mark=x, mark size=2, mark options={solid}]
table {%
64 7.21125983503166
128 7.51228983069564
256 7.81331982635962
512 8.1143498220236
1024 8.41537981768758
};
\addplot [ultra thick, dashed, red, mark=o, mark size=2, mark options={solid}]
table {%
64 7.21874446244101
128 7.51983334697984
256 7.8208927840871
512 8.1219374997243
1024 8.4229748551878
};
\addlegendentry{Lang-Firsov}
\addlegendentry{Vaccum State}
\end{axis}

\end{tikzpicture}
    \caption{\raggedright $T$-gate cost, on a Logarithmic scale against the number of sites, $N$. We show the costs of preparing either the vacuum state with $U_0$ and the Lang-Firsov state with the added cost of implementing controlled $\tilde{U}_{\alpha}$'s. In both cases, we set $m=6$ (encoding phononic degrees of freedom at a site). We see that the difference in the cost of state preparation circuits are barely visible in this scale. }
    \label{fig:tcountfullcircuit}
\end{figure}

The costs of QPE will scale with the costs of preparing oracle access to $e^{iH}$, either with a Trotter decomposition as was done in \cite{Macridin_PRA}, or a more sophisticated technique such as a linear combination of Pauli strings (LPC) \cite{Zhang_2024}. These are similar to the costs of our algorithm, so the clearer demonstration of the value of our algorithm comes from comparing the costs of our state preparation against the cost of preparing the vacuum state. As shown in Fig.~\ref{fig:tcountfullcircuit}, the cost between preparing the vacuum state, with $U_0$, is almost undistinguishable from the cost of preparing the Lang-Firsov state, with $\tilde{U}_{\alpha}$. 

The costs of a full QPE protocol, agnostic to the costs of each QPE step, will scale with the number of times it is repeated, which is $\mathcal{O}(\Omega_{\text{GS}}^{-1})$. So, a fair comparison between the costs of our method and the costs of sing a vacuum state is to look at the ratio
\begin{equation}
    \frac{\Omega^{-1}_{\text{GS}}(\text{LF}) \; T_{\text{LF}}}{\Omega^{-1}_{\text{GS}}(\text{vac})\;  T_{\text{vac}}}.\label{eq:cost_ration}
\end{equation}

As shown in Fig.~\ref{fig:ratio-tcount-allqpesteps}, costs of our protocol scale much more favorably than the vacuum state, because the overlap with the initial state does not decay exponentially. Note that the number of sites does not strong change the ratio. This is because, as seen in Fig.~\ref{fig1}, the number of sites does not strongly affect the overlaps $\Omega$.

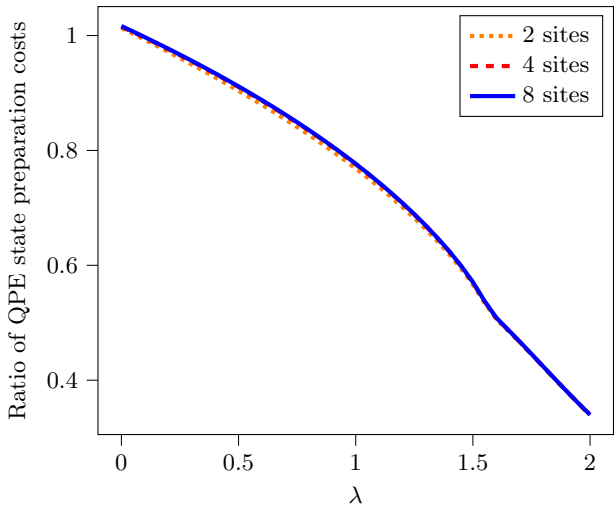
\begin{figure}[t]
    \centering
\begin{tikzpicture}

\definecolor{darkgray176}{RGB}{176,176,176}
\definecolor{darkorange25512714}{RGB}{255,127,14}
\definecolor{forestgreen4416044}{RGB}{44,160,44}
\definecolor{steelblue31119180}{RGB}{31,119,180}

\begin{axis}[
tick align=outside,
tick pos=left,
x grid style={darkgray176},
xlabel={$\lambda$},
xmin=-0.1, xmax=2.1,
xtick style={color=black},
y grid style={darkgray176},
ylabel={Ratio of QPE state preparation costs },
ymin=0.305374813748262, ymax=1.05042132531815,
ytick style={color=black}
]
\addplot [ultra thick, orange, dotted]
table {%
0 1.0132444598338
0.05 1.00304440885969
0.1 0.992704060285302
0.15 0.982215338885572
0.2 0.971569523168177
0.25 0.960757177566075
0.3 0.949768075893632
0.35 0.938591114022036
0.4 0.927214210140461
0.45 0.915624189648219
0.5 0.903806651292532
0.55 0.891745810000083
0.6 0.879424310472569
0.65 0.866823003765559
0.7 0.853920676649156
0.75 0.840693720238026
0.8 0.827115719805134
0.85 0.813156941190953
0.9 0.798783681299077
0.95 0.783957432755592
1 0.768633795711454
1.05 0.752761028391599
1.1 0.736278064942536
1.15 0.719111706700204
1.2 0.701172456695804
1.25 0.68234797268863
1.3 0.662492022308397
1.35 0.641404192060966
1.4 0.618788543542222
1.45 0.594157712007209
1.5 0.566574851767821
1.55 0.534246823450288
1.6 0.506210518629378
1.65 0.486263523061371
1.7 0.466076520549391
1.75 0.445209477729302
1.8 0.423959449696964
1.85 0.402641743954531
1.9 0.381521958168985
1.95 0.360812417913893
2 0.34067951763477
};
\addplot [ultra thick, red, dashed]
table {%
0 1.01545186980609
0.05 1.00585730728068
0.1 0.996087826348571
0.15 0.986134920465949
0.2 0.975989428318091
0.25 0.965641466417432
0.3 0.955080353202458
0.35 0.944294522741718
0.4 0.933271426442732
0.45 0.921997419951482
0.5 0.910457631873653
0.55 0.898635809934261
0.6 0.886514138436776
0.65 0.874073019333288
0.7 0.861290806558297
0.75 0.848143479264102
0.8 0.834604236450435
0.85 0.820642993363506
0.9 0.806225708062041
0.95 0.791313603607683
1 0.775862055149537
1.05 0.759819035531658
1.1 0.743123223414539
1.15 0.725701473510886
1.2 0.707464687823643
1.25 0.688299478471591
1.3 0.668060416752015
1.35 0.646546732568588
1.4 0.623461534329197
1.45 0.598312738417839
1.5 0.570149118702255
1.55 0.537108420383891
1.6 0.508142661729752
1.65 0.487403020732697
1.7 0.466555846312279
1.75 0.445137751043877
1.8 0.423438587380399
1.85 0.401769686234496
1.9 0.380388110545319
1.95 0.359488148417889
2 0.339240564274166
};
\addplot [ultra thick, blue]
table {%
0 1.01655557479224
0.05 1.00696349741
0.1 0.997194683989268
0.15 0.987240795499527
0.2 0.977092645659939
0.25 0.966740483279585
0.3 0.956173379226867
0.35 0.945380729658783
0.4 0.934348949815248
0.45 0.92306518443673
0.5 0.911514445179148
0.55 0.899680232722346
0.6 0.887545405715252
0.65 0.875091023009286
0.7 0.862291845500273
0.75 0.84912818019913
0.8 0.835571412229714
0.85 0.821592242438423
0.9 0.807156591857236
0.95 0.792226685597173
1 0.776755458366965
1.05 0.760687346799546
1.1 0.74397108447385
1.15 0.726526993386557
1.2 0.708269347930912
1.25 0.689077515100144
1.3 0.668815353089634
1.35 0.647281447459317
1.4 0.624161386391556
1.45 0.598999383610212
1.5 0.570797649326451
1.55 0.537709987391981
1.6 0.508731295224355
1.65 0.48795918556916
1.7 0.467090753720897
1.75 0.445680632949496
1.8 0.423954763401153
1.85 0.402280197981044
1.9 0.380881794920829
1.95 0.359978398236032
2 0.339744580715051
};
\addlegendentry{$2$ sites}
\addlegendentry{$4$ sites}
\addlegendentry{$8$ sites}
\end{axis}

\end{tikzpicture}
    \caption{ \raggedright We compare the T-count cost of QPE-based ground-state preparation using two different initial states. For fixed $\omega_0 = 1$, the figure shows the cost ratio~\cref{eq:cost_ration} as a function of the coupling constant dependent parameter $\lambda$. As $\lambda$ increases, the relative cost of the LF-based protocol decreases rapidly, with only weak dependence on the system size $N$.}
    \label{fig:ratio-tcount-allqpesteps}
\end{figure}

\section{Discussion and Outlook}
\label{sec:conclusion}

To conclude, we have argued that using as the initial state the Lang-Firsov ansatz of Eq. (\ref{ini}) with a finite value of $\alpha$ (instead of the no-phonon, $\alpha=0$ initial state proposed before) can very significantly increase the overlap between this initial state and the true GS, and thus lower significantly the needed number of QPE iterations. The cost of implementing the Lang-Firsov ansatz has been calculated and the difference between the no-phonon initial state $\alpha=0$ and the much better finite $\alpha$ options turns out to be very low. 

While this work offers proof-of-principle for the advantage of starting with physically-motivated, better initial states when studying systems with electron-phonon coupling, the reality is that the properties of the  single Holstein polaron have already been calculated to extremely high accuracy with multiple bias-free classical computational methods \cite{FehskeTrugman2007}. The same holds for many other single polaron models in the linear coupling approximation and for Einstein (dispersionless) phonons, such as the SSH/Peierls model \cite{Dominic}, the lattice Fr\"ohlich model \cite{AlexandrovKornilovitch1999}, the breathing-mode coupling \cite{Bayo2007}, etc. More recently there has also been progress in investigating polarons in models with non-linear electron-phonon couplings and Einstein phonons, with similar methods \cite{Li2015,Ragni2023,Zhang2023}. A much less studied problem is the formation of single polarons due to coupling to acoustic (gapless) phonons \cite{Hahn2021}, and we are not aware of any numerical results for polarons in models which include both acoustic and optical phonons. Perhaps this is a niche where quantum computations might become competitive, especially if other realistic ingredients such as different types of couplings to different types of phonons, are included. We note that for all such single polaron models, one can find simple variational guesses similar to Eq. (\ref{ini}), or deploy the more involved but much more accurate Momentum Average variational approximation \cite{BerciuPRL06,Dominic,BerciuPRL13,Dominic17}. Both can be extended to the study of bipolarons (two polarons bound through exchange of bosons between their clouds) \cite{John}, as well as coupling of carriers to other interesting bosons, such as magnons in a magnetically ordered background \cite{Bayo,Hadi}.

Another related problem  where quantum computation advantage might emerge is for Hamiltonians with electron-phonon coupling at finite but very low electron concentrations. This limit is difficult for classical computational schemes especially in higher dimensions, because it requires modeling very large clusters in order that the number of considered electrons be large enough to smooth out number fluctuations. A  few approximation-free numerical results exist \cite{Mishchenko2005,Mishchenko2014,Mishchenko2021,Johnston2022,Johnston2023} but they are far from a complete exploration of the entire parameter space for all  electron-phonon coupling models of interest. For these problems, a good guess of the initial state for a quantum computation, especially at stronger couplings,  could be obtained from the variational Momentum Average approximation, which  has recently also been generalized to the low carrier concentration regime \cite{Berciu2022,Nocera2023}. 

Finally, we conclude by noting that realistic simulations of any material, be it a quantum material (a solid-state crystal in the thermodynamic limit) or a molecule relevant for biological or medicinal purposes, will require dealing with both electron-electron interactions and electron-boson couplings, on equal footing. From this perspective, progress in proposing initial states that better encode electron-electron correlations must go hand-in-hand with improvements in the description of the bosonic clouds accompanying these quasiparticles. The work presented here is a first step in this direction.

\vspace{5pt}

\section{Acknowledgments}

A.A., M.B. and S.E.S.S. are indebted to Stepan Fomichev for the initial motivation for this project and for suggestions regarding PennyLane. A.A. thanks Rio Weil for insightful discussions on circuit decompositions. This project was undertaken thanks in part
to funding from the Max Planck-UBC-UTokyo Center for
Quantum Materials and the Canada First Research Excellence
Fund, 
as well as the Natural Sciences and Engineering Research
Council of Canada (A.N. and M. B.). A.A. is supported by the Humbolt Foundation. S.E.S.S is funded by Quantum Valley Lower Saxony and also acknowledges the support of the Natural Sciences and Engineering Research Council of Canada (NSERC), PGS D - 587455 - 2024.

Cette recherche a été financée par le Conseil de recherches en sciences naturelles et en génie du Canada (CRSNG), PGS D - 587455 - 2024.

\section*{Data Availability}
Our code is available at \url{https://github.com/Skeltonse/warm-start}.
\bibliography{literature}

\appendix
\onecolumngrid
\section{Pre-processing for the state preparation circuit}
Our state preparation method for $\ket{vac}_B$ is based on \cite{gilyenqsvt2019}. Given a circuit preparing a block-encoding of 
\begin{equation}
    A=\sum_{x=\frac{-N}{2}}^{\frac{N}{2}}\sin\left(\frac{2x}{N}\right)\ket{x}\bra{x},
\end{equation}
one can use a variety of quantum signal processing based templates \cite{dongefficient2021, gilyenqsvt2019} to prepare a block-encoding of  $\sum_{x}\frac{f(x')}{\left|\left|f\right|\right|_{\infty}}\ket{x}\bra{x}$, $U_{\sin}$, where $\left|\left|f\right|\right|_{\infty}$ is the function norm over the appropriate region $x'\in[a, b]$\footnote{
For the shifted gaussian $g_{\alpha}(x)$, we would instead use a block-encoding of 
$\sum_{x}\sin\left(\frac{x}{N}\right)\ket{x}\bra{x}$. An early pre-print version of \cite{mcardle2025quantumstatepreparationcoherent}
 gives a circuit for the $(1, 1, 0)$ block-encoding of this.}. 

\subsection{Shifted Gaussians}
Beginning from desired function $f:\left[a, b\right]\rightarrow\mathbb{R}$, we use a polynomial approximation to
\begin{equation}
    h(y):=f\left((b-a)\arcsin(y)+a\right).\label{eq:qsvtvariableshiftfcn}
\end{equation}
where $y\in\left[0, \sin(1)\right]$ as in \cite{mcardle2025quantumstatepreparationcoherent}. 

In our case, the desired function is 
\begin{equation}
    g_{\alpha}(x)=e^{\frac{-(x-\alpha)^2}{2}}.\label{eq:shiftedguassianfcn}
\end{equation}
where $x\in\left[-W, W\right]$.

Using \ref{eq:qsvtvariableshiftfcn} and \ref{eq:shiftedguassianfcn}, we find that the QET polynomial is, written in variable $z=\arcsin(y)$, $z\in[0, 1]$,
\begin{equation}
    h_{\alpha}(z):=e^{-{2W^2\left(z-\frac{W+\alpha}{2W}\right)^2}}.
\end{equation}
Finally, we need to check the normalization of $f$, because QET requires $\left|\left|f\right|\right|\leq 1$. We simply have
\begin{align}
h_{\alpha}'(z)&=-4W^2\left(z-\frac{W+\alpha}{2W}\right)e^{-{2W^2\left(z-\frac{W+\alpha}{2W}\right)^2}}\\
h_{\alpha}'(0)&\rightarrow z=\frac{W+\alpha}{2W}, h_{\alpha}(1)=1
\end{align}
so then \cref{eq:qsvtapprox} is the function which must be entered into the phase-factor finding algorithm. 


In the highly-relevant case $\alpha=0$, we can use that $g_{\alpha}$ is even to simplify the problem. We begin from
\begin{equation}
    h(y):=f\left(2a\arcsin(y)\right).\label{eq:qsvtvariableshiftfcnsymmetric}
\end{equation}
where again $y\in\left[0, \sin(1)\right]$, leading to
\begin{equation}
    h_{0}(z):=e^{-\frac{W^2}{2}z^2}.\label{eq:qsvtapprox}
\end{equation}
qsppack can easily generate $p_d(x)$, a degree-d polynomial approximation of $h_0(y)$.  Instead of using Chebyshev-expansion bounds to determine the degree, we numerically identify a $\mathcal{O}(10^{-2})$ even-parity approximation, see \cref{fig:polynomial-function-error}. We then generate a symmetric-QET angle set with negligible error (\cref{fig:polynomial-qsp-error}). If a higher-precision approximation is required, then one can use a polynomial approximation without parity restrictions. 

\begin{figure}
    \centering
  \input{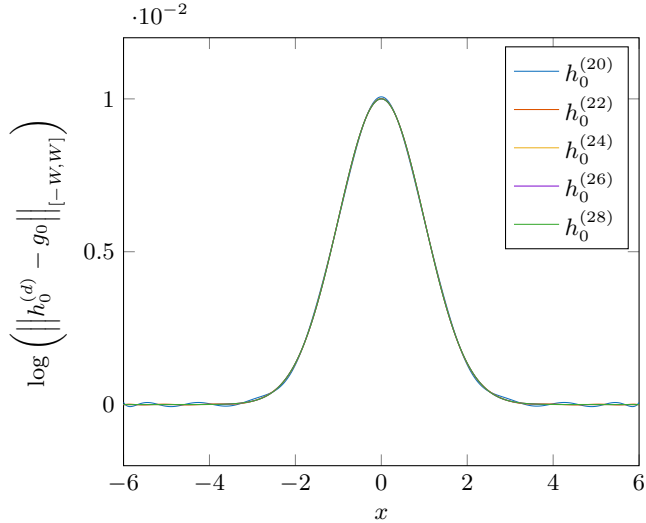}
    \caption{Function norm error between $e^{x^2/2}$ and different $h_0(z)$ approximations for degrees $\{20, 22, 24, 26, 28\}$ shown on a log scale. We observe that the error spikes close to $x=0$, and that there is not much benefit to slightly increasing the degree after $d={22}$. }
    \label{fig:polynomial-function-error}
\end{figure}

\begin{figure}
    \centering
    \input{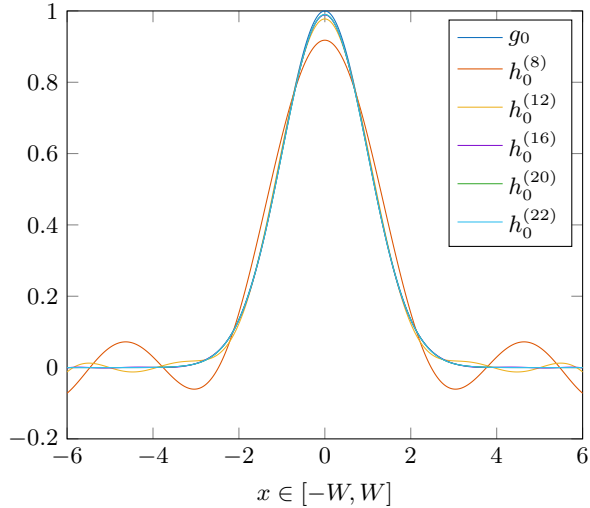}
    \caption{Even polynomial approximations of $h_0(z)$ for degrees $d\in[8, 22]$. We see that the polynomials converge to $e^{-x^2/2}$ as the degree increases.}
    \label{fig:polynomial-qsp-error}
\end{figure}

\subsection{Pre-processing and algorithm templates}
Generally, there is a trade-off between how simple a QET circuit will be and the generality of the polynomials which can be implemented with the template. 

Our state preparation circuits use 'symmetric qsp' (symmetric QET in the multisignal case)  \cite{dongefficient2021}. This template has the strongest phase-factor finding results \cite{alexis2024infinitequantumsignalprocessing, alexis2024quantumsignalprocessingnonlinear, ni2024fastphasefactorfinding}, and the angles it computes can be straightforwardly used in QET or quantum singular value transformation circuits. Our polynomial approximation is real with definite parity, so the angles  can easily be computed using "qsppack"\cite{qsppackcitation}.

$U_{QET}$ has costs which are fully determined by the cost of the block-encoding and the highest polynomial degree. The two QET routines are then added using a linear combination of unitaries on the circuit, completing 
the state preparation unitary.

\subsection{Hilbert Spaces and Registers}
We consider a system of \( N \) lattice sites \( j = 0, \dots, N-1 \), 
containing a single electron and one truncated local phonon mode per site.

\paragraph{Electron register \( E \).}
In the first-quantized representation, the electron position is stored in a 
register spanned by basis states labelled by the site index \( j \in \{0, \dots, N-1\} \).
The Hilbert space of the electron register is
\begin{equation}
    \mathcal{H}_E
    = \operatorname{span}\bigl\{\, \lvert j \rangle_E : j = 0,\dots,N-1 \,\bigr\}
    \subset (\mathbb{C}^2)^{\otimes k_E},
    \qquad k_E = \lceil \log_2 N \rceil .
\end{equation}
Each basis state \(\lvert j \rangle_E\) corresponds to the binary encoding of the site label,
\begin{equation}
    \lvert j \rangle_E
    = \lvert b_{k_E-1} b_{k_E-2} \dots b_0 \rangle,
    \qquad
    j = \sum_{\ell=0}^{k_E-1} b_\ell\, 2^\ell, 
    \quad b_\ell \in \{0,1\}.
\end{equation}
A general electron state is therefore
\begin{equation}
    \lvert \psi_E \rangle
    = \sum_{j=0}^{N-1} c_j\, \lvert j \rangle_E,
    \qquad 
    \lvert \psi_E \rangle \in \mathcal{H}_E .
\end{equation}

\paragraph{Bosonic registers \( B_j \).}

Each lattice site \( j \) hosts a local phonon mode whose infinite-dimensional Fock space
is truncated to a finite dimension \( d_B = n_{\max} + 1 \).
We choose \( m \) such that \( 2^m \ge d_B \), and associate to site \( j \) an \(m\)-qubit register \(B_j\).

\subparagraph*{(i) Qubit-level representation.}
Initially, each bosonic register is prepared in a fiducial computational state 
\(\lvert 0 \rangle^{\otimes m}_{B_j}\),
which serves as the hardware vacuum of the register.
This state carries no physical meaning yet---it merely represents the all-zero configuration
of the qubits before loading the truncated phonon state.
Collectively, the fiducial vacuum of all bosonic registers is
\begin{equation}
    \lvert 0 \rangle_{B}^{\otimes N}
    = \bigotimes_{j=0}^{N-1} \lvert 0 \rangle^{\otimes m}_{B_j}.
\end{equation}

\subparagraph*{(ii) Physical phonon Hilbert space.}
The physical phonon degrees of freedom are encoded in the truncated Fock basis
\(\{\,\lvert \nu_j \rangle_{B_j}\,:\,\nu_j=0,\dots,d_B-1\,\}\),
which span the logical Hilbert space of register \(B_j\):
\begin{equation}
    \mathcal{H}_{B_j}
    = \operatorname{span}\bigl\{\, \lvert \nu_j \rangle_{B_j} : \nu_j = 0,\dots,d_B - 1 \,\bigr\}
    \subset (\mathbb{C}^2)^{\otimes m}.
\end{equation}
The global bosonic Hilbert space is then
\begin{equation}
    \mathcal{H}_B = \bigotimes_{j=0}^{N-1} \mathcal{H}_{B_j},
    \qquad 
    \dim(\mathcal{H}_B) = d_B^{\,N}.
\end{equation}
A generic bosonic basis state is written as
\begin{equation}
    \lvert \nu_0, \ldots, \nu_{N-1} \rangle_B
    = \bigotimes_{j=0}^{N-1} \lvert \nu_j \rangle_{B_j}.
\end{equation}

\subparagraph*{(iii) Initialization and state preparation.}
In the simulation, one first initializes all registers in the fiducial unphysical state
\(\lvert 0 \rangle_{B}^{\otimes N}\)
and subsequently applies a sequence of unitaries that map this state into
the desired phonon configuration within the truncated space,

\paragraph{Full system state.}
The total Hilbert space of the system is given by
\begin{equation}
    \mathcal{H} = \mathcal{H}_E \otimes \mathcal{H}_B,
\end{equation}
and a general pure state can be expanded as
\begin{equation}
    \lvert \Psi \rangle
    = \sum_{j=0}^{N-1} c_j\, \lvert j \rangle_E
    \bigotimes
    \sum_{\nu_0, \ldots, \nu_{N-1} = 0}^{d_B - 1}
        c^{(j)}_{\nu_0, \ldots, \nu_{N-1}}
        \lvert \nu_0, \ldots, \nu_{N-1} \rangle_B .
\end{equation}

\paragraph{Qubit count.}
The number of logical qubits required  to put this system on a quantum computer as it is is given by
\begin{equation}
    Q_{\mathrm{total}} 
    = k_E + N m 
    = \lceil \log_2 N \rceil + N m .
\end{equation}

\paragraph{Interpretation of the gates.}

The operator \(U_0\) denotes a state-preparation unitary acting on the bosonic register \(B_i\),
which consists of \(m\) qubits encoding the truncated local phonon Hilbert space.
It maps the fiducial computational basis state of that register,
\(|0^{\otimes m}\rangle_{B_i}\),
to the physical phonon vacuum,
\begin{equation}
U_0\,|0^{\otimes m}\rangle_{B_i}
= |\nu = 0\rangle_{B_i}
\;\approx\;
\frac{1}{\mathcal{N}}
\sum_{x_j \in \mathcal{G}}
e^{-\tfrac{x_j^2}{2\sigma^2}}\,|x_j\rangle_{B_i},
\end{equation}
That is, \(U_0\) prepares on the \(m\)-qubit register \(B_i\) a discretized Gaussian wavefunction
approximating the ground state of the local harmonic oscillator in the position basis.
The sum runs over the discrete grid points \(x_j \in [-W, W]\)
represented on those \(m\) qubits, and \(\mathcal{N}\) some suitable normalization constant.
In this form, the register encodes the oscillator vacuum
\(|\nu=0\rangle_{B_i}\) as a superposition of position eigenstates
\(|x_j\rangle_{B_i}\) with Gaussian amplitudes \(e^{-x_j^2/(2\sigma^2)}\),
where \(\sigma = \sqrt{\hbar/(m_{\mathrm{ph}}\omega_0)}\) sets the width of the ground-state wavepacket.

The controlled operation \(\tilde{U}_{\alpha}\) implements a local phonon displacement
conditioned on the electron being located at site \(i\).
We define
\begin{equation}
    \tilde{U}_{\alpha} \; \equiv \; U_{\alpha}\,U_{0}^{\dagger},
\end{equation}
so that it maps the local phonon ground state, previously prepared by \(U_0\),
to the corresponding displaced (coherent) state.
Indeed,
\begin{equation}
    \tilde{U}_{\alpha}\,|\nu = 0\rangle_{B_i}
    \equiv U_{\alpha}\,U_{0}^{\dagger}\,|\nu = 0\rangle_{B_i}
    = U_{\alpha}\,|0^{\otimes m}\rangle_{B_i}
    =: |\alpha\rangle_{B_i}.
\end{equation}
Here \(U_{\alpha}\) is the state-preparation unitary that, when acting on the fiducial
computational state \(|0^{\otimes m}\rangle_{B_i}\), prepares a discretized coherent state
with displacement amplitude \(\alpha\) on those \(m\) qubits.
In the position-basis representation this corresponds to translating the Gaussian
wavefunction by a distance \(\alpha\) such that
\begin{equation}
    U_{\alpha}\,|0^{\otimes m}\rangle_{B_i}
    \approx
    \frac{1}{\mathcal{N}}
    \sum_{x_j} e^{-\tfrac{(x_j - \alpha. \sigma)^2}{2\sigma^2}}\,|x_j\rangle_{B_i}
    = |\alpha\rangle_{B_i}.
\end{equation}
Hence, \(\tilde{U}_{\alpha}\) acts as the qubit-encoded representation of the
continuous-variable displacement operator
\begin{equation}
    \hat D_{B_i}(\alpha) = e^\frac{\left({\alpha b_i^\dagger - \alpha^* b_i} \right)}{\sqrt{2}} ,
\end{equation}
which satisfies \(\hat D_{B_i}(\alpha)\,|\nu=0\rangle_{B_i} = |\alpha\rangle_{B_i}\).
Thus, the sequence \(U_0\) followed by the controlled application of \(\tilde{U}_{\alpha}\)
first prepares each bosonic register in the ground state of its local oscillator
and then conditionally displaces the phonon cloud associated with the electron’s site
by a distance \(\alpha\),
while all other phonon modes remain in their vacuum states
\(|\nu = 0\rangle_{B_j}\) for \(j \neq i\).






\subsection{Implementing the displacment operator via QFT--Diagonal--iQFT scheme}

For a real displacement parameter \(\alpha\in\mathbb{R}\), the displacement operator reduces to a position translation,
\[
D(\alpha)=e^{\alpha a^\dagger-\alpha a}=e^{-i x_0 P},\qquad x_0=\sqrt{2}\,\alpha  \; .
\] Since translations in position are generated by \(P\), the unitary \(e^{-i x_0 P}\) is \emph{diagonal in the momentum basis} with eigenvalues \(e^{-i x_0 p}\). This motivates the construction: move to momentum space (QFT), apply the diagonal phase, and move back (inverse QFT).

\paragraph{Discrete encoding and QFT conventions.}
An \(m\)-qubit register (\(N=2^m\) points) encodes position on the half-open interval \([ -W,\, W )\) with spacing
\[
\Delta x=\frac{2W}{N},\qquad x_j=-W+j\,\Delta x,\quad j=0,\dots,N-1,
\]
and momentum with spacing
\[
\Delta p=\frac{2\pi}{N\,\Delta x}=\frac{\pi}{W},\qquad
p_k=\Big(k-\frac{N}{2}\Big)\Delta p,\quad k=0,\dots,N-1.
\]
The unitary QFT relates the bases as
\[
|p_k\rangle=\frac{1}{\sqrt N}\sum_{j=0}^{N-1}e^{+i\frac{2\pi}{N}jk}|x_j\rangle,\qquad
|x_j\rangle=\frac{1}{\sqrt N}\sum_{k=0}^{N-1}e^{-i\frac{2\pi}{N}jk}|p_k\rangle.
\]
The grid-encoded HO vacuum is
\[
|\psi_{\mathrm{vac}}\rangle=\frac{1}{\mathcal N}\sum_{j=0}^{N-1}\exp\!\Big(-\frac{x_j^2}{2\sigma^2}\Big)\,|x_j\rangle,
\]
with \(W\) and \(\Delta x\) chosen so the Gaussian is well resolved and negligible at the boundaries.

\paragraph{Circuit}
We realize the displacement as
\[
U(x_0):=e^{-i x_0 \tilde P}
=\mathrm{iQFT}\,\Big[\mathrm{diag}\big(e^{-i x_0 p_k}\big)\Big]\mathrm{QFT}.
\]
\begin{figure}
    \centering
  \begin{tikzpicture}
    \begin{yquant}
        qubits {$\ket{j}$} j[1];
qubit {$\ket{vac}_B$} b[6];

box {QFT} (b);
box {$R_z\left(\phi_{0}\right)$} b[0]| j;
box {$R_z\left(\phi_{1}\right)$} b[1]|  j;
box {$R_z\left(\phi_{2}\right)$} b[2]| j;
box {$R_z\left(\phi_{3}\right)$} b[3]|  j;
box {$R_z\left(\phi_{4}\right)$} b[4] | j;
box {$R_z\left(\phi_{5}\right)$} b[5] | j;
box {$\text{QFT}^{\dag}$} (b);
    \end{yquant}
\end{tikzpicture}
    \caption{$\tilde{U}_{\alpha}$, controlled on some $j$ used to implement the displacement of the phonon-cloud required for our Lang-Firsov ansatz. }
    \label{fig:utildecircuit}
\end{figure}
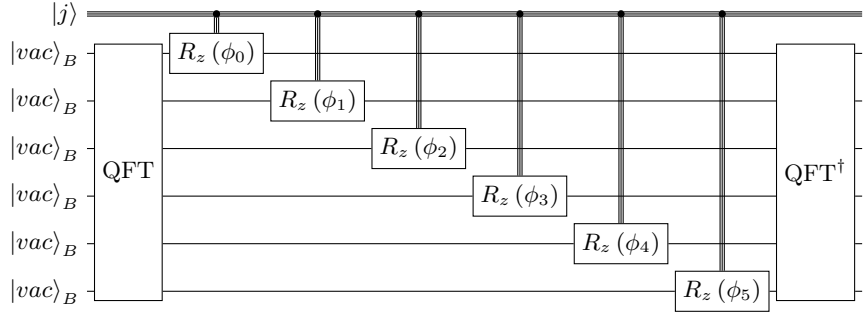
\emph{Gate-level description :}
\begin{enumerate}
\item Apply \textbf{QFT} on the \(m\)-qubit register (maps \(|x\rangle\to|p\rangle\)).
\item Apply the \textbf{momentum-diagonal phase}. Writing the (uncentered) index \(k=\sum_{\ell=0}^{m-1} b_\ell 2^\ell\) with \(b_\ell\in\{0,1\}\), and dropping the global phase from the \(-N/2\) shift,
\[
e^{-i x_0 p_k}=e^{-i x_0 \Delta p\,k}
=\prod_{\ell=0}^{m-1}\exp\!\big(-i\,b_\ell\,\phi_\ell\big),
\qquad
\boxed{\ \phi_\ell=x_0\,\Delta p\,2^\ell=\sqrt{2}\,\alpha\,\frac{\pi}{W}\,2^\ell\ }.
\]
Operationally: apply single-qubit \(R_z(\phi_\ell)\) to qubit \(\ell=0,\dots,m-1\) (all in parallel).
\item Apply \textbf{iQFT} on the same \(m\) qubits (maps \(|p\rangle\to|x\rangle\)).
\end{enumerate}

\paragraph{Result and correctness.}
By construction,
\[
U(x_0)=\sum_{k=0}^{N-1} e^{-i x_0 p_k}\,|p_k\rangle\langle p_k|
\]
is diagonal in momentum with eigenphase \(-x_0 p_k\), i.e., the finite-grid realization of \(e^{-i x_0 P}\). In the position basis,
\[
\langle x_j|U(x_0)|x_{j'}\rangle
=\frac{1}{N}\sum_{k=0}^{N-1}e^{+i\frac{2\pi}{N}k\,(j'-j-s)},\qquad s=\frac{x_0}{\Delta x},
\]
the inverse DFT of a linear phase (Dirichlet kernel). If \(s\in\mathbb{Z}\), this is an exact cyclic shift by \(s\) sites; for general \(s\), it implements the sampled continuous translation by \(x_0\). Therefore,
\[
U(x_0)\,|\psi_{\mathrm{vac}}\rangle
\approx \frac{1}{\mathcal N}\sum_{j=0}^{N-1}\exp\!\Big(-\frac{(x_j-x_0)^2}{2\sigma^2}\Big)\,|x_j\rangle
= D(\alpha)\,|\psi_{\mathrm{vac}}\rangle,\qquad x_0=\sqrt{2}\alpha,
\]
up to exponentially small wrap-around error when the packet is well contained in \([ -W,\, W )\).

\subsection{block-encodings}



We need to prepare a block-encoding of 
\begin{equation}
   \sum_{x=-N/2}^{N/2-1}\sin\left(\frac{2x}{N}\right)\ket{x+\frac{N}{2}}\bra{x+\frac{N}{2}}= \sum_{y=0}^{N-1}\sin\left(\frac{2y}{N}-1\right)\ket{y}\bra{y}
\end{equation}
Where $y=x+\frac{N}{2}$, but since $\ket{y}, \ket{x}$ are just state vectors labels we can switch them out without confusion. In binary form, each term in the sine function can be written
\begin{equation}
     \sum_{y=0}^{N-1}y_{j}2^{j+1-m} -1\cdot 2^1.
\end{equation}
The circuit \Cref{fig:be-symm-circuit} works by encoding $R_y(b_{m-1}2^{-1}+b_{m-2}2^{-2}+...b_02^{-m})$ where each $b_j\in\{0,1\}$  is determined by the controlled gates. The result,  $U_{\sin}(N)$, is a $(1, 1, 0)$ block-encoding for  $   \sum_{x=-N/2}^{N/2-1}\sin\left(\frac{2x}{N}\right)\ket{x+\frac{N}{2}}\bra{x+\frac{N}{2}}$, meaning that it exactly prepares $   \sum_{x=-N/2}^{N/2-1}\sin\left(\frac{2x}{N}\right)\ket{x+\frac{N}{2}}\bra{x+\frac{N}{2}}$ with one additional qubit and no subnormalization.

Alternatively, we could have implemented a signed circuit as is done in \cite{mcardle2025quantumstatepreparationcoherent}.
\begin{figure}
\begin{tikzpicture}
\begin{yquant}
qubit {$\ket{a_{\idx}}$} a[1];
qubit {$\ket{y_{\idx}}$} y[3];

box {$R_y\left(-1\right)$} a;
box {$R_y\left(2^{1-m}\right)$} a| y[0];
box {$R_y\left(2^{2-m}\right)$} a|  y[1];
box {$R_y\left(2^{0}\right)$} a | y[2];
X a;
\end{yquant}
\end{tikzpicture}
\caption{Circuit implementation of  $\sum_{x=-N/2}^{N/2-1}\sin\left(\frac{2x}{N}\right)\ket{x+\frac{N}{2}}\bra{x+\frac{N}{2}}$ in a block-encoding. Note that here, we use $R_y(\theta)=e^{-i\theta Y}$. }
\label{fig:be-symm-circuit}
\end{figure}
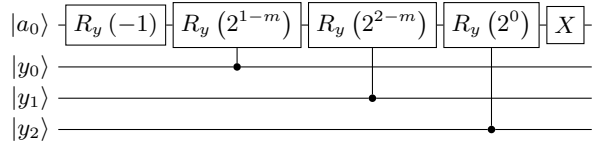
\subsection{QET Circuits}
Each $U_0$ will be implemented with a real, definite parity approximation and so we can use symmetric-QSP circuits \cite{dongefficient2021}. The block-encoding used, \cref{fig:be-symm-circuit}, is not Hermitian, and so we begin from the quantum eigenvalue transformation circuit with a non-Hermitian block-encoding given in \cite{dongefficient2021}.

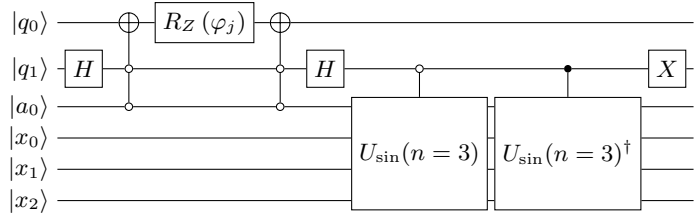
\begin{figure}
\begin{tikzpicture}
\begin{yquant}
qubit {$\ket{q_{\idx}}$} q[2];
qubit {$\ket{a_{\idx}}$} a[1];
qubit {$\ket{x_{\idx}}$} x[3];

box {$H$} q[1];
not q[0] ~ q[1], a;
box {$R_Z\left(\varphi_j\right)$} q[0];
not q[0] ~ q[1], a;
box {$H$} q[1];
box {$U_{\sin}(n=3)$} (x, a) ~ q[1];
box {$U_{\sin}(n=3)^{\dag}$} (x, a) | q[1];
box {$X$} q[1];
\end{yquant}
\end{tikzpicture}
\caption{One step of the symmetric QSP circuit, $U_{QSP}(\phi_j)$, using the convention that $R_Z=e^{-i\phi Y}$.}
\label{fig:qsp-nonhermit-step}
\end{figure}

\begin{figure}
\begin{tikzpicture}
\begin{yquant}
qubit {$\ket{q_{\idx}}=\ket{0}$} q[2];
qubit {$\ket{a_{\idx}}=\ket{0}$} a[1];
qubit {$\ket{x_{\idx}}$} x[3];

box {$H$} q[1];
box {$U_{QSP}(\phi_j)$} (q, a, x);
inspect {$...$} (q, a, x);
box {$H$} q[1];
not q[0] ~ q[1], a;
box {$R_Z\left(-\varphi_0\right)$} q[0];
not q[0] ~ q[1], a;
dmeter q[1], a;
\end{yquant}
\end{tikzpicture}
\caption{Symmetric-QSP circuit implementing a complex, definite-parity, degree $d$ polynomial of a block-encoded Hamiltonian, where the oracle is a non-Hermitian block-encoding $U_{\sin}$. Usually this circuit will be modified to implement a LCU step, preparing a real polynomial instead of a complex one.}
\label{fig:qsp-nonhermit-circuit}
\end{figure}
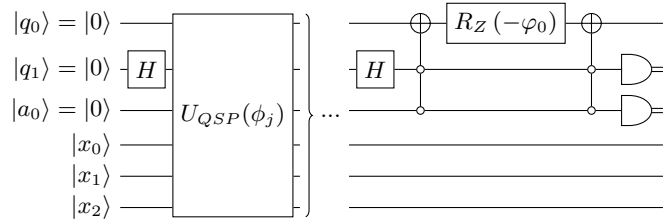

\Cref{fig:qsp-nonhermit-circuit} encodes a complex polynomial $P$ and we are interested in only the real part. We can prepare $P^*(x)$ in QSP and then use a linear combination of unitaries to encode $\frac{1}{2}\left(P(x)+P^*(x)\right)$. 
The LCU is extremely simple,
\begin{align}
    H\otimes I \cdot C_0U_{QSP}(\Phi) \cdot C_1U_{QSP}(-\Phi) \cdot H\otimes I.
\end{align}
where $-\Phi$ defines a QSP circuit encoding the complex conjugate of $P$. Now, we review why the $1$-qubit reduction in LCU step of real-symmetric QSP with Hermitian block-encodings also works in the case of non-Hermitian block-encodings.

The phase factors of the QSP circuit for $P^*(x)$ are derived from the phase factors of $P(x)$, and we have \begin{equation}
    -\Phi=\{\varphi_0+\pi, \varphi_1+\pi, ....\varphi_{d-1}+\pi, \varphi_{d}\}
\end{equation}
using the $\varphi$ convention, which can be derived from the symmetric-QSP convention, sets $\{\phi_j\}$ using
\begin{equation}
   \varphi_j= \begin{cases}
        \phi_0+\frac{\pi}{4}, & j=0\\
        \phi_j+\frac{\pi}{2}, & 1\leq j\leq d-1\\
        \phi_n+\frac{\pi}{4}, & j=d.
    \end{cases}
\end{equation}
In this text, $\Phi, -\Phi$ always denote sets of $\{\varphi_j\}$ and not $\{\phi_j\}$.
The conversion between $\Phi, -\Phi$ allows one to simplify the LCU step. \Cref{fig:qsp-nonhermit-circuit} is already encoding a QSP circuit with $-\varphi_j$ factors in the $\ket{1}$ branch of the state. We need to encode an additional $+\pi$ phase on the $\ket{1}$ branch of the state in the LCU step, which can be done with a single $Z$ gate. So, we simply add $H$ gates and $Z$ gates on all but the $\phi_d$ iteration. The final circuit is given in \cref{fig:qsp-nonhermit-circuit-real}.
\begin{figure}
\scalebox{0.75}[0.75]{
\begin{tikzpicture}
\begin{yquant}
qubit {$\ket{q_{\idx}}=\ket{0}$} q[2];
qubit {$\ket{a_{\idx}}=\ket{0}$} a[1];
qubit {$\ket{x_{\idx}}$} x[3];

box {$H$} q;

box {$H$} q[1];
not q[0] ~ q[1], a;
box {$R_Z\left(\varphi_d\right)$} q[0];
not q[0] ~ q[1], a;
box {$H$} q[1];
box {$U_{\sin}$} (x, a) ~ q[1];
box {$U_{\sin}^{\dag}$} (x, a) | q[1];
box {$X$} q[1];

inspect {$...$} (q, a, x);

box {$H$} q[1];
box {$Z$} q[0];
not q[0] ~ q[1], a;
box {$R_Z\left(\varphi_j\right)$} q[0];
not q[0] ~ q[1], a;
box {$H$} q[1];
box {$U_{\sin}$} (x, a) ~ q[1];
box {$U_{\sin}^{\dag}$} (x, a) | q[1];
box {$X$} q[1];

inspect {$...$} (q, a, x);

box {$H$} q[1];
box {$Z$} q[0];
not q[0] ~ q[1], a;
box {$R_Z\left(-\varphi_0\right)$} q[0];
not q[0] ~ q[1], a;

box {$H$} q;
dmeter q[1], a;
\end{yquant}
\end{tikzpicture}
}
\caption{Symmetric-QSP circuit implementing a real, definite-parity, degree $d$ polynomial where the oracle is a npn-Hermitian block-encoding $U_{\sin}$}
\label{fig:qsp-nonhermit-circuit-real}
\end{figure}
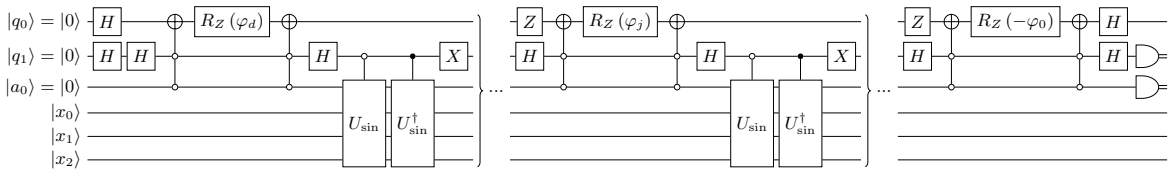
\subsection{Exact Amplitude Amplification}\label{sec:qaa}
We use exact amplitude amplification for QET circuits as given in \cite{mcardle2025quantumstatepreparationcoherent}. The success probability is given by 
\begin{equation}
    p_{succ}=\left(\frac{1}{2}\frac{\sqrt{\sum_{x=0}^{N-1}\left|p_d\right|^2}}{\sqrt{N\left|\left|p_d\right|\right|^2_{\max}}}\right)^2
\end{equation}
where $\left|\left|p_d\right|\right|^2_{\max}$ is the maximum value of $p_d$ over the interval.

Numerically, we find that the success probability stabilizes to a value independent of $M$ after a small number of qubits, $\sqrt{p_{succ}}=0.0369$ leading to
\begin{equation}
    m'=\lceil\frac{\pi}{4\cdot \arcsin(\sqrt{p_{succ}})}-\frac{1}{2}\rceil=4.
\end{equation}
The cost of the full state preparation scheme will scale with $m'$. Because QAA is a common subroutine, and our implementation very closely follows \cite{mcardle2025quantumstatepreparationcoherent}, we do not review it further herein.

\end{document}